\documentclass[conference]{IEEEtran}

\usepackage{amsfonts}
\usepackage{setspace}

\usepackage{setspace}
\usepackage{amssymb}
\usepackage{epsfig}
\usepackage{graphicx}
\usepackage{xcolor}
\usepackage{algorithm}
\usepackage{algorithmicx}
\usepackage{algpseudocode}
\usepackage{amsmath}

\floatname{algorithm}{Algorithm}

\usepackage[ps2pdf,
bookmarks=false,
bookmarksnumbered=false, % true means bookmarks in
bookmarksopen=false, % true means only level 1
colorlinks=false]{}

\begin{document}

\title{Actor-Critic Deep Reinforcement Learning for Dynamic Multichannel Access}

\author{Chen Zhong, Ziyang Lu, M. Cenk Gursoy, and Senem Velipasalar
\\Department of Electrical Engineering and Computer Science,
\\Syracuse University, Syracuse, NY 13244
\\Email: czhong03@syr.edu, zlu112@syr.edu, mcgursoy@syr.edu, svelipas@syr.edu}

\maketitle

\begin{abstract}
We consider the dynamic multichannel access problem, which can be formulated as a partially observable Markov decision process (POMDP). We first propose a model-free actor-critic deep reinforcement learning based framework to explore the sensing policy. To evaluate the performance of the proposed sensing policy and the framework's tolerance against uncertainty, we test the framework in scenarios with different channel switching patterns and consider different switching probabilities. Then, we consider a time-varying environment to identify the adaptive ability of the proposed framework. Additionally, we provide comparisons with the Deep-Q network (DQN) based framework proposed in \cite{wang2018deep}, in terms of both average reward and the time efficiency.
\end{abstract}


\section{Introduction}
Recent exponential growth in mobile data traffic coupled with the scarce and underutilized nature of spectral resources has increased the importance of efficient dynamic spectrum access strategies. Dynamic spectrum access is challenging especially if the channel conditions in different frequency bands vary over time and such variations and their patterns are not known a priori. Motivated by these, we in this work propose a deep reinforcement learning based framework for dynamic multichannel access.

We consider an environment with $N$ correlated channels, and each channel has two possible states: good or bad. And we assume that the channel switching can be modeled as a Markov chain with at most $2^N$ states. And the target is to select the good channels as many times as possible. Since the environment is unknown to users, the user can only try sensing different channels at each time and determine the pattern as much as possible based on its own observation.

Therefore, the core of this problem is to find a dynamic sensing policy for the unknown environment based on the user's individual information from channels after each time of sensing. Taking advantage of the ability of reinforcement learning methods in exploring unknown environments \cite{Peters2005NaturalA}\cite{sutton1998reinforcement}, we design a deep reinforcement learning framework, and use it as an agent to interact with the multichannel access environment and dynamically make sensing decisions. Considering its benefits in terms of storage and time efficiency, we choose the natural actor-critic algorithm, since this algorithm requires no replay memory and the update only depends on the current and next observation, which reduces the computational complexity as well.

Our main contributions in this work can be summarized as follows:
\begin{itemize}
	\item We propose an actor-critic algorithm based deep reinforcement learning framework for the dynamic multichannel access problem, and this framework can work with relatively larger number of channels than other previous deep reinforcement learning based frameworks.
	\item We analyze the performance of the proposed framework and compare it with the deep Q-network (DQN) framework presented in \cite{wang2018deep}. Simulation results demonstrate that our proposed framework can achieve competitive performance for the case of $16$ channels, and better performance for the cases of $32$ and $64$ channels.
	\item We test the proposed approach in time-varying scenarios, and the results show the adaptive ability of actor-critic deep reinforcement learning. Also, our framework leads to significant benefits in terms of time efficiency.
\end{itemize}

\section{Related Work}

The multichannel access problem has been widely studied in the literature. For the correlated channel scenarios, the authors in \cite{zhao2007decentralized} developed an analytical framework for opportunistic spectrum access based on the theory of partially observable Markov decision processes (POMDP). And for independent channels, the problem can be modeled as a restless multi-armed bandit process (RMAB) \cite{liu2008restless}.

To solve the opportunistic spectrum access problem, in which the user needs to probe the channel before sensing, previous studies focused on seeking effective sensing policies. For instance, myopic policies were studied in \cite{zhao2008myopic} \cite{ahmad2009optimality}, where the belief information of channels are collected through sufficient statistics and the user only senses the channel with the highest conditional probability. A stochastic game theory based policy was presented in \cite{xu2012opportunistic} \cite{zheng2015stochastic}, where multiple users are in the system but each user can adjust its behavior based on the individual information. In \cite{wang2017optimally}, a joint probing and accessing policy was proposed to allow the user to probe multiple channels at a time.

In recent years, reinforcement learning has attracted much interest to seek solutions for more practical and complicated multichannel access problems. Authors in \cite{dai2014online} proposed a continuous sampling and exploitation (CSE) online learning algorithm for a RMAB model. In \cite{zhang2014model}, Q-learning was applied in the sensing order selection problem, and it was shown that Q-learning can be used in the presence of imperfect sensing. More recently, a DQN based framework for the PODMP model was proposed in \cite{wang2018deep}, where the performance is analyzed in different scenarios. However, limited by the learning method, the aforementioned works are currently not able to handle a large number of channels. To tackle this challenge, our work presents an actor-critic framework that can operate in scenarios with a large number of channels.

\section{System Model}

In this work, we consider the dynamic multichannel access problem where the user dynamically selects one out of $N$ channels. Each channel has two possible states: the good channel state, which allows the user to transmit successfully, and the bad channel state, which will lead to transmission failure. We assume that before sensing the channel, the user does not know the states of the channels.  And it is also not efficient for a user to sense all channels every time. Therefore, we assume that the user senses a single channel at a time and learns its state. Also, we assume that the state of each channel is dynamically switching between good and bad, which is modeled as a Markov chain. Hence, for the user, choosing a channel at a good state out of $N$ channels, whose states are varying, is a POMDP, in which the user aims to learn the pattern of variations of channel states based on previous decisions.

To solve this pattern recognition problem, we propose an actor critic algorithm based deep reinforcement learning framework. In this framework, an agent will observe the environment, get the feedback from the channels, and update the decision policy. As we mentioned, the environment has $N$ channels, and the switching pattern is unknown to the agent. We assume that, at time $t$, the channel state can be denoted as $\mathtt{X}_t = \{\mathtt{x}_1, \mathtt{x}_2, ..., \mathtt{x}_{N} \} $, where $N $ is the total number of channels, $\mathtt{x}_i$ stands for the state of the $i^{th}$ channel. For every channel $i$, $i = 1, 2, ..., N$, we have $\mathtt{x}_i = 1$ if the channel is in good state, or $\mathtt{x}_i = 0$ if the channel is in bad state. And each time the agent senses a channel, the state can be either good or bad. Therefore, we define the reward as follows: if a good channel is chosen, the reward $r_t$ will be $+1$; otherwise, the reward $r_t$ will be $-1$.

The agent's observation can be denoted as $O_t = \{o_1, o_2, ..., o_N \}$, where $N$ is the total number of channels. If channel $i$, $i = 1, 2, ..., N$, is chosen, the agent senses it and knows its channel state, so we define $o_i = r_t$; otherwise, the agent will record $o_i = 0$. The agent will learn on the basis of its previous experience. We assume the agent keeps an observation space $\mathcal{O}$ that consists of the most recent $M$ observations. The observation space is initialized as an all-zero $N \times M$ matrix, and at each time $t$, the latest observation $O_t$ will be added to the observation space, and oldest observation $O_{t-M}$ will be removed. The updated observation space $\mathcal{O}$ for time $t+1$ can be denoted as $\mathcal{O}_{t+1} = \{O_{t}; O_{t-1}; ...; O_{t-(M-1)} \}$.

Next, we consider a discrete action space $\mathcal{A} = \{1, 2, ..., N \}$, where $N$ is the total number of channels. Each valid action in the action space describes the index of the channel that will be sensed. Hence, when an action is chosen, the agent will sense the corresponding channel and receive the reward which reveals the condition of the chosen channel. The agent can only choose one channel to sense/learn at each iteration. The aim of the agent is to find a policy $\pi$, which maps the observation space $\mathcal{O}$ to the action space $\mathcal{A}$,  that maximizes the long-term expected reward $R$ of channel sensing decisions:
\begin{equation*}
%\underset{\pi^*}{\text{maximize}} \quad R
\pi^* = \arg \max_{\pi} R
\end{equation*}
where $\pi^*$ denotes the optimal decision policy, and in a finite time duration $T$, we express $R$ as :
\begin{equation*}
R = \frac{1}{T} \sum_{i = 1}^{T} r_i
\end{equation*}
And according to the definition of $R$, we have $R \in [-1, 1]$.

\section{Actor Critic Framework}\label{sec:framework}
In this section, we describe the proposed actor critic algorithm based framework for multichannel access.

\subsection{Algorithm Overview}\label{sub:algorithm}
In this part, we introduce the actor critic algorithm. The actor critic structure consists of two neural networks: actor and critic. In our work, the two networks will not share any neurons but both of the two networks are parameterized by $\theta$.

\emph{Actor:} The actor is employed to explore a policy $\pi$, that maps the agent's observation $\mathcal{O}$ to the action space $\mathbb{R}^a$:
\begin{equation*}
\pi_{\theta}(\mathcal{O}) : \mathcal{O} \rightarrow \mathbb{R}^a
\end{equation*}
So the mapping policy $\pi_{\theta}(\mathcal{O})$ is a function of the observation $\mathcal{O}$ and is parameterized by $\theta$. And the chosen actor can be denoted as
\begin{equation*}
a = \pi_{\theta}(\mathcal{O})
\end{equation*}
where we have $a \in \mathcal{A}$. In our work, since the action space is discrete, we use softmax function at the output layer of actor network, so that we can obtain the scores of each actions. The scores are summed to $1$, and stand for the probabilities to obtain a good reward by choosing the corresponding actions.

\emph{Critic:} The critic is employed to estimate the value function $V(\mathcal{O})$. At time instant $t$, when the action $a_t$ is chosen by the actor network, the agent will execute it in the environment and send the current observation $\mathcal{O}_t$ along with the feedback from the environment to the critic. The feedback includes the reward $r_t$ and the next time instant observation $\mathcal{O}_{t+1}$. Then, the critic can calculate the TD (Temporal Difference) error:
\begin{equation*}
\delta^{\pi_\theta} = r_t + \gamma V(\mathcal{O}_{t+1}) - V(\mathcal{O}_t)
\end{equation*}
where $\gamma \in (0,1)$ is the discount factor.

\emph{Update:}The critic is updated by minimizing the least squares temporal difference (LSTD):
\begin{equation*}
%\underset{V^*}{\text{minimize}} \quad (\delta^{\pi_\theta} )^2
V^* = \arg \min_{V} (\delta^{\pi_\theta} )^2
\end{equation*}
where $V^*$ denotes the optimal value function.

The actor is updated by policy gradient. Here we use TD error to compute the policy gradient:
\begin{equation*}
\Delta_{\theta} J(\theta) = E_{\pi_{\theta} } [ \nabla_{\theta} \log \pi_\theta(\mathcal{O}, a)  \delta^{\pi_\theta} ]
\end{equation*}
where $\pi_\theta(\mathcal{O}, a)$ denotes the score of action $a$ under the current policy.
As introduced before, the policy is actually to score each of the actions based on current observation. And since we use softmax at the output layer of the actor network, the scores of actions should be $\pi_\theta(\mathcal{O}, a) \in [0,1] $. Hence, we have $\log \pi_\theta(\mathcal{O}, a) \le 0 $, and thus, to seek for the gradient decent direction, the actor must be updated as
\begin{equation*}
%\underset{\pi^*}{\text{maximize}} \quad \nabla_{\theta} \log \pi_\theta(\mathcal{O}_t, a_t)  \delta^{\pi_\theta}
\pi^* = \arg \max_{\pi}\nabla_{\theta} \log \pi_\theta(\mathcal{O}_t, a_t)  \delta^{\pi_\theta}
\end{equation*}
where $\pi^*$ denotes the optimal actor policy.

The full framework is shown in Algorithm \ref{alg:AC} below.

\subsection{Related Definitions}
In this part, we will introduce some related definitions in the framework.

\emph{Channel State and Agent's Observation:} The channel state is varying as described by a Markov chain and it's a part of the environment, which is is unknown to the agent. So, the agent can only take its own observation space $\mathcal{O}$ as the input of the actor critic framework. Based on the POMDP, the agent can only sense the chosen channel in every iteration, and observes the reward which depends on the chosen channel's state. As defined in the previous section, the observation space is a sparse matrix with only one nonzero element in each column.

\emph{Action:} The actor network will score all possible actions in the action space $\mathcal{A}$, and the action with the highest score will be chosen. In our setting, the action means to sense the corresponding channel.

\emph{Reward:} The reward is received when the action is executed, which means that the agent chooses a channel and gets direct feedback from the environment. The reward is defined based on the condition of the chosen channel. Since each of the channels has two possible states: good or bad, the agent will receive $+1$ if the chosen channel is in good state, or $-1$ if the chosen channel is in bad state.

\begin{algorithm}
	\caption{Actor-Critic Deep Reinforcement Learning Algorithm for Dynamic Multichannel Access}
	\label{alg:AC}
	\begin{algorithmic}
		\State Initialize the critic network $V_{\theta}(\mathcal{O} )$ and the actor $\pi_{\theta}(\mathcal{O})$, parameterized by $\theta$.
		\State The environment initializes the state of each channel $\mathtt{X}$.
		\State The agent initializes its observation as all zero matrix $\mathcal{O}_0$		
		\For{$t = 0,T$}
		
		\State With the observation, the agent selects an action $a_t = \pi(\mathcal{O}_t| \theta)$ w.r.t. the current policy
		\State Agent senses the chosen channel and receives the reward $r_t$ based on the channel state.
		
		\State Critic calculates the TD error: $ \delta^{\pi_\theta} = r_t + \gamma V(\mathcal{O}_{t+1}) - V(\mathcal{O}_t) $
		\State Update the critic by minimizing the loss: $\mathcal{L}(\theta) = (\delta^{\pi_\theta} )^2$
		\State Update the actor policy by maximizing the action value: $\Delta\theta = \nabla_{\theta} \log \pi_\theta(\mathcal{O}_t, a_t)  \delta^{\pi_\theta}$
		
		\State Update the network parameters:
		\State \qquad\qquad\qquad
		$\theta' \longleftarrow \tau \theta' + (1-\tau)\theta'$	
		\State Update the observation $\mathcal{O}$.	
		\State Update the channel state $\mathtt{X}$.
		
		\EndFor
	\end{algorithmic}
\end{algorithm}

\section{Experiments}
In this section, we present the simulation setting of our proposed actor critic framework, and evaluate the performance through experiments and provide comparisons with the DQN based framework proposed in \cite{wang2018deep}.
\subsection{Simulation Setup}
The agent consists of two networks: actor and critic. Each of the two networks has two layers. For the actor, which scores all actions in the action space, the first layer has $200$ neurons with ReLU as the activation function, and second layer has $N$ neurons with Softmax as the activation function, where $N$ is the total number of channels. For the critic, which calculates the value of the chosen action, the first layer has $200$ neurons with ReLU as the activation function, and second layer has $1$ neuron. Especially, since the critic will evaluate the decision made by the actor, the learning rate of the critic network should be larger than that of the actor network. Here, we set the learning rate of the critic network as $0.0005$, and the learning rate of the actor network as $0.0001$. To ensure stability, both learning rates will decay exponentially with the decay rate as $0.95$ for every $5000$ time slots.

In the environment, we denote the Markov chain of channel states as $\mathcal{P}$, which defines the switching probabilities between states.

\subsection{Results}
In this section, we present the experiment results, and compare our framework with the DQN framework.

\emph{DQN Framework} proposed in \cite{wang2018deep} consists of two hidden layers, and maintain a replay memory with the size of $1,000,000$. To update the network, the DQN framework will replay a minibatch of $32$ samples extracted from the memory.

\subsubsection{Single Good Channel}
\begin{figure}
	\centering
	 \includegraphics[width=1\linewidth]{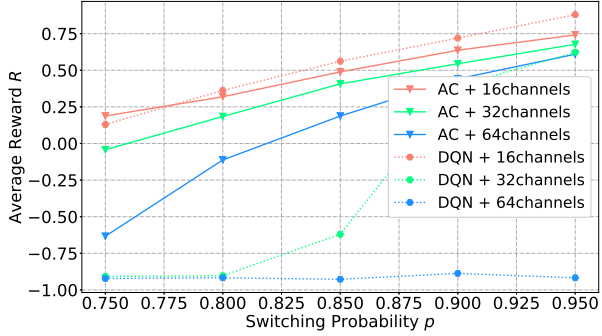}
	\caption{Average reward vs. switching probability. We consider $16$, $32$, $64$ channels cases with the switching probability varies as $p = \{0.75, 0.80, 0.85, 0.90, 0.95\}$}
	\label{fig:roundrobin1}
\end{figure}

In this experiment, we consider the total number of channels $N = \{16, 32, 64\}$, and only one channel is good for every time slot. To evaluate the performance, we calculate the expected reward $R$ with different  Markov chains $\mathcal{P}$. To define a Markov Chain for the channel distribution, we need to specify the channel states in order and the state switching probabilities. We assume that, for each state, the probability that current state will transfer to another state is $p$, and the probability that the current state will be kept is $1-p$. Our experiments were conducted in two cases:

\textbf{Round-Robin Switching Scenario:}
In this experiment, we assume that the index of the only  good channel switches from $1$ to $N$ according to a round-robin scheduling. And we vary the switching probabilities as $p = \{0.75, 0.80, 0.85, 0.90, 0.95\}$. We compare the actor-critic (AC) policy with DQN in terms of the average reward. As shown in Fig. \ref{fig:roundrobin1}, the DQN framework cannot handle the case of $64$ channels for any switching probability; and for the case of 32 channels, the DQN framework achieves negative rewards when $p \le 0.85$. For the actor-critic framework, though there is also negative reward achieved with 64 channels at $p \le 0.80$ and with $32$ channels at $p = 0.75$, the reward is greater than that of DQN in the same setting, which means that our proposed actor-critic framework is more suitable when the number of channels is relatively large. As to the overall tendency, except for the \lq\lq DQN $+\ 64$channels\rq\rq \ curve, we can observe increasing average reward as the switching probability increases and the gap between different cases diminishes in the meantime. According to the definition, $p$ stands for the probability of switching between states, which means higher switching probability will decrease the uncertainty and will make it easier for the agent to learn the policy. Comparing the performance at $p = 0.75$, the actor-critic framework demonstrates benefits for all $N$ values, and therefore it has higher tolerance against uncertainty.
%\begin{figure}
%\centering
%\includegraphics[width=1.0\linewidth]{roundRobin}
%\caption{Average reward vs. switching probability. We consider $16$, $32$, $64$ channels cases with the switching probability varies as $p = \{0.75, 0.80, 0.85, 0.90, 0.95\}$}
%\label{fig:roundRobin}
%\end{figure}

\textbf{Arbitrary Switching Scenario:}
In the round-robin switching scenario, the channel states switch according to a specific scheduling model. However, this information is unknown to the actor-critic agent, and of course is not being used in the process to find a sensing policy. Moreover, the actor-critic algorithm was proposed as a model-free algorithm. To demonstrate the performance of the proposed framework in a model-free environment, we in this experiment, fix the switching probability $p$ at $0.9$ and test the framework with $10$ different arbitrary switching orders.

\begin{figure}
\centering
\includegraphics[width=\linewidth]{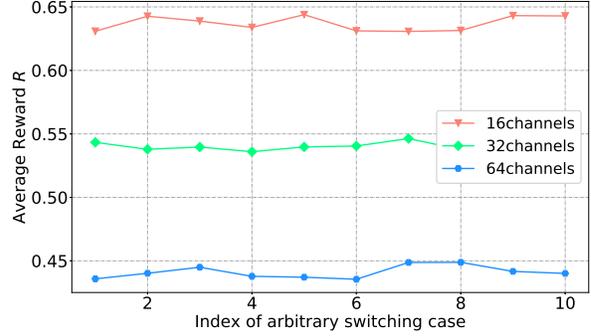}
\caption{The average reward for different arbitrary switching orders}
\label{fig:arbitrary}
\end{figure}

Fig. \ref{fig:arbitrary} plots the performance in the cases of $16$, $32$ and $64$ channels with $10$ randomly generated arbitrary switching orders. For any given number of channels, the average reward varies only slightly across different switching cases, showing that our proposed framework can work in a model-free environment.

%\subsubsection{Multiple Good Channels Situation}
%Perfectly correlated scenario

\subsubsection{Time-Varying Environment}
As discussed before, both the proposed actor-critic framework and the DQN framework introduced in \cite{wang2018deep} are reward-driven algorithms which can continually interact with the environment and update the policies. To illustrate the adaptive ability of the proposed framework, we have designed a time-varying environment, where at the beginning, the agent has been trained for pattern $\mathcal{P}_1$, and at time slot $t = 500$, the channel distribution changes to the second pattern $\mathcal{P}_2$, but the change point is unknown to the agent. The experiment was conducted with a fixed switching probability $p = 0.9$, and arbitrary switching order where $32$ channels are grouped into $8$ subsets randomly and each subset has $4$ perfectly correlated channels.

\begin{figure}
	\centering
	 \includegraphics[width=\linewidth]{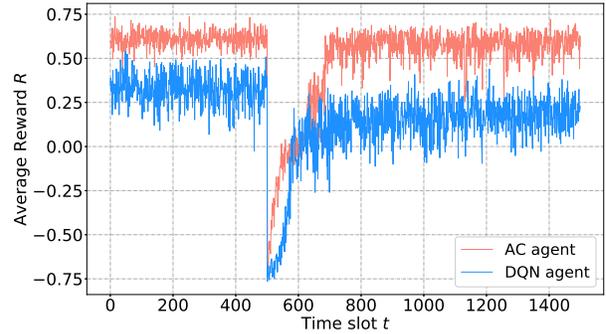}
	\caption{The re-training process in a time-varying environment with the change point at $t = 500$.}
	\label{fig:timevarying}
\end{figure}

The re-training process is shown in Fig. \ref{fig:timevarying} in terms of the reward averaged over every $500$ sensing decisions. When the channel state switching pattern changes at $t = 500$, the average reward achieved by both actor-critic framework and DQN drops to negative values suddenly. And then, over time, the policies get updated and adapted to the new pattern, so the average rewards gradually increase and reach to the previous levels. Comparing the time duration it takes for the agent to get back the previous level, we conclude that our proposed framework is very competitive in terms of the adaptive ability, though the actor-critic structure which has two separate neural networks takes slightly more time to converge.

\subsubsection{Study of Runtime}
To meet the real-time requirements, the sensing decision must be made quickly. To highlight the efficiency of the actor-critic framework, we have computed the average runtime needed for making one decision and compared it with that needed in the DQN framework. Table \ref{my-label} shows the runtime for one decision needed by the actor-critic (AC) agent and the DQN agent for the case of having a single good channel out of $N$ channels in total, where $N = \{16, 32, 64\} $.

\begin{table}[]
	\centering
	\caption{The runtime need for every sensing decision}
	\label{my-label}
	\begin{tabular}{|c|c|c|c|l}
		\cline{1-4}
		number of channels & AC agent & DQN agent & \% reduced &  \\ \cline{1-4}
		16                 & 0.002428 & 0.025381  & 90.4328    &  \\ \cline{1-4}
		32                 & 0.003998 & 0.030833  & 87.0340    &  \\ \cline{1-4}
		64                 & 0.004002 & 0.059308  & 93.2527    &  \\ \cline{1-4}
	\end{tabular}
\end{table}

The proposed actor-critic framework is actually more complicated in architecture because it has two neural networks and hence has more parameters to update. But we only pass one actor to the critic, so that the critic requires less computational resources. Another important reason why our framework can have significant savings in the runtime is that we do not need to replay any experience because the LSTD of the critic network is enough to ensure that the actor policy is updating in the correct direction, while the DQN proposed in \cite{wang2018deep} replayed $32$ samples for each time of updating to make the sensing policy stable. For the current number of channels and users, the second reason for the substantial improvements in runtime is that the action space is limited. But once that action space increases, as the number of channels increases, the first reason will become more significant.

%\begin{table}[]
%	\centering
%	\caption{My caption}
%	\label{my-label}
%	\begin{tabular}{|c|c|c|c|l}
%		\cline{1-4}
%		number of channels & AC agent              & DQN agent            & \% reduced        &  \\ \cline{1-4}
%		16                 & 0.002428196353646134  & 0.025380523323043365 & 90.4328357507052  &  \\ \cline{1-4}
%		32                 & 0.0039978530936624936 & 0.03083336821744746  & 87.03400463592473 &  \\ \cline{1-4}
%		64                 & 0.004001687273381366  & 0.059308025571986195 & 93.25270528771145 &  \\ \cline{1-4}
%	\end{tabular}
%\end{table}

%\subsubsection{Multi-user}
%
%To make equal comparison, we fix the number of channels at $N = 16$ or $32$ (depends on the results) and the transfer probability as $p = 0.9$. The number of user varies as $U = 2, 3, 4$.
%
%Currently, we are not able to allow more users because the increase in user number will cause very large action space. It's hard to converge. And, the sparse observation space also makes the frame work hard to converge.

\section{Conclusion}

In this work, we have considered the dynamic multichannel access problem modeled as a POMDP. To effectively find the sensing policy, we have proposed and implemented a model-free actor-critic deep reinforcement learning framework. Then, we have tested the framework on round-robin and arbitrary switching scenarios, and compared the average reward with that of the DQN framework. We have demonstrated the proposed framework's ability in handling a larger number of channels and high tolerance against uncertainty. To highlight the adaptive ability, we have conducted simulations in a time-varying environment. The proposed framework has shown competitive performance with respect to DQN in terms of time slots needed for the re-training phase. Finally, we have demonstrated the efficiency of the actor-critic framework by computing the percentage of runtime that can be saved compared to the DQN framework.

For future work, we are interested in making the proposed framework more practical in more complicated environments like the system with very large number of channels and multiple users.

\bibliographystyle{IEEEtran}
\bibliography{Channel_selection}

\end{document}